\documentclass[aps,prd,preprintnumbers,showpacs,nofootinbib,groupedaddress,superscriptaddress,11pt]{revtex4-2}

\usepackage{amsmath}
\usepackage{amsfonts}
\usepackage{amssymb}
\usepackage{bm}
\usepackage{mathtools}
\usepackage{mathrsfs}
\usepackage{url}
\usepackage{hyperref}
\usepackage{xcolor}
\definecolor{darkgreen}{rgb}{0.0, 0.3, 0.0} 
\definecolor{darkviolet}{rgb}{0.3, 0.1, 0.6} 
\hypersetup{
    colorlinks=true,       
    linkcolor=blue,        
    citecolor=darkgreen,   
    urlcolor=darkviolet    
}
\makeatletter

\@addtoreset{equation}{section}
\makeatother

\usepackage[normalem]{ulem}

\colorlet{RED}{red}

\colorlet{green}{green!60!black}

\begin{document}

\title{General relativistic effects on photon spectrum emitted from dark matter halos around primordial black holes}

\author{Toya Suzuki}
\email{toya.suzuki@grad.nao.ac.jp}
\affiliation{Division of Science, National Astronomical Observatory of Japan, 2-21-1 Osawa, Mitaka, Tokyo 181-8588, Japan}
\affiliation{School of Physical Sciences, Graduate University for Advanced Studies (SOKENDAI), 2-21-1 Osawa, Mitaka, Tokyo 181-8588, Japan}

\author{Takahisa Igata}
\email{takahisa.igata@gakushuin.ac.jp}
\affiliation{Department of Physics, Gakushuin University, Mejiro, Toshima, Tokyo 171-8588, Japan}

\author{Kazunori Kohri}
\email{kazunori.kohri@gmail.com}
\affiliation{Division of Science, National Astronomical Observatory of Japan, 2-21-1 Osawa, Mitaka, Tokyo 181-8588, Japan}
\affiliation{School of Physical Sciences, Graduate University for Advanced Studies (SOKENDAI), 2-21-1 Osawa, Mitaka, Tokyo 181-8588, Japan}
\affiliation{Theory Center, IPNS, and QUP (WPI), KEK, 1-1 Oho, Tsukuba, Ibaraki 305-0801, Japan}
\affiliation{Kavli IPMU (WPI), UTIAS, The University of Tokyo, Kashiwa, Chiba 277-8583, Japan}

\author{Tomohiro Harada}
\email{harada@rikkyo.ac.jp}
\affiliation{Department of Physics, Rikkyo University, Toshima, Tokyo 171-8501, Japan}

\date{\today}

\begin{abstract}
We investigate general relativistic effects on the photon spectrum emitted from decaying (or annihilating) particle dark matter in the halo surrounding a primordial black hole. The spectrum undergoes significant modification due to gravitational redshifts, which induces broadening as a result of the intense gravitational field near the black hole. This characteristic alteration in the photon spectrum presents a unique observational signature. Future observations of such spectral features may provide critical evidence for a mixed dark matter scenario, involving both primordial black holes
and particle dark matter.
\end{abstract}


\maketitle

\newpage

\section{Introduction}
Dark matter has been estimated to be approximately five times more abundant than baryonic (or visible)
matter~\cite{Planck:2018vyg}, making the elucidation of its fundamental nature a central research topic in astronomy, astrophysics, cosmology, gravitational physics,
and particle physics. In this context, the term ``dark matter" 
refers specifically to Cold Dark Matter (CDM). 
Some of the proposed candidates for dark matter, 
e.g.,
Weakly Interacting Massive Particles, Axion-Like Particles,
and right-handed neutrinos, 
possess the potential to emit electromagnetic radiation in the present Universe through annihilation or decay 
(see, e.g., Refs.~\cite{Jungman:1995df,Bertone:2004pz,Bershady:1990sw,Masso:1995tw,Asaka:2005an,Abazajian:2006yn}
and references therein). Consequently, their identification can be feasible through observational studies utilizing electromagnetic wave telescopes.

Meanwhile, the possibility that primordial black holes (PBHs) constitute a component of dark matter has been the subject of significant recent debate~\cite{Carr:2009jm,Carr:2020gox,Carr:2020xqk,Green:2020jor,Escriva:2022duf,Byrnes:2025tji}. A PBH is a black hole 
that formed in the early Universe~\cite{Carr:1974nx,Carr:1975qj}. It is well established that if small-scale density fluctuations are sufficiently large, gravitational collapse can occur in the early Universe, leading to the formation of PBHs~\cite{Carr:1975qj,Carr:1994ar,Harada:2013epa}. This is a 
unique scenario for 
black hole formation
that does not involve the process of a star's death. In this study, we primarily consider PBHs with masses in the range of approximately $10^{17}$--$10^{23}$~g~\cite{Carr:2009jm,Carr:2020gox}, which are capable of comprising 100$\%$ of dark matter (see also 
review articles~\cite{Carr:2020xqk,Green:2020jor,Escriva:2022duf}, and references therein). This mass range has attracted considerable attention, as PBHs within this regime can account for the total dark matter content without conflicting with any known observational constraints.

In addition, even in scenarios where PBHs do not constitute the entirety of dark matter, their presence can induce notable cosmological phenomena. In this work, we refer to non-PBH dark matter candidates, such as those mentioned above, as 
``particle dark matter",
and we investigate a mixed dark matter scenario that incorporates both PBHs and particle dark matter. An intriguing aspect of this scenario is that particle dark matter candidates are expected to form a halo around a PBH due to structure formation processes~\cite{Mack:2006gz,Ricotti:2007au}. 
This results in the formation of a distinctive dark matter halo that does not emerge in the standard scenarios involving only particle dark matter. Notably, the density profile of such a halo follows a special feature of a power-law distribution, e.g., $\rho(r) \propto r^{-9/4}$, which has 
been studied only using Newtonian mechanics ~\cite{Bertschinger:1985wig,Berezinsky:2013fxa,Adamek:2019gns,Serpico:2020ehh}. A PBH enveloped by this characteristic halo, referred to as a 
``dressed PBH,"
is expected to produce electromagnetic emissions due to the decay or annihilation of surrounding particle dark matter. Moreover, a distinctive emission spectrum should be anticipated in the vicinity of the PBH, significantly redshifted as a result of gravitational effects predicted by general relativity.

In this paper, we conduct a precise calculation of the modifications to photon emissions from dark matter surrounding dressed PBHs due to gravitational redshift effects based on the full general relativistic calculations. The detection of such characteristic spectral features would provide deeper insights into the physical properties of dressed PBHs, a phenomenon inherently associated with the mixed dark matter scenario comprising both PBHs and particle dark matter. Furthermore, future observational advancements in electromagnetic wave detection hold the potential to unveil crucial information about the true nature of dark matter.

Throughout this paper, we adopt a unit system in which $\hbar = c = k_{\mathrm{B}}
= 1$.

\label{sec:intro}

\section{Notes on Dark-Matter Dressed Primordial Black Holes in Newtonian Mechanics}
According to the pioneering works of
Refs.~\cite{Mack:2006gz,Ricotti:2007au}, the mass of 
a halo composed of particle dark matter, such as WIMPs, around a PBH grew on the order of $M_{\mathrm{halo}} \propto (1 + z)^{-1}$ at
redshift $z$ due to the structure formation after the matter-radiation equality $z < z_{\rm eq} \sim 3400$.

Then, the explicit evolution of
$M_{\mathrm{halo}}$ is analytically estimated to be
\begin{equation}
    \label{eq:2:1}
    M_{\mathrm{halo}} = 3 \Big( \frac{1000}{1+z} \Big) M_{\mathrm{PBH}}.
\end{equation}
In this case, the outer radius of the halo is given by
\begin{equation}
    \label{eq:2:2}
    r_{\mathrm{max}} = 0.019 \,\mathrm{pc} \Big( \frac{M_{\mathrm{halo}}}{M_{\odot}} \Big)^{1/3} \Big( \frac{1000}{1+z} \Big),
\end{equation}
which remains valid until $z \sim 30$, the redshift at which standard halo formation begins in the $\Lambda$CDM model.
Here,
$M_{\mathrm{PBH}}$ represents the mass of a PBH, and
$M_{\odot}$ denotes the solar mass.
It was also shown that, during the radiation-dominated
epoch until the matter-radiation equality,
$M_{\mathrm{halo}}$ had grown to a level comparable to the mass of a PBH,
$M_{\mathrm{halo}} \sim {\cal O}(1)
M_{\mathrm{PBH}}$~\cite{Mack:2006gz,Ricotti:2007au}.
Thus, we can use Eq.~(\ref{eq:2:1}) just after the matter-radiation equality $z \le z_{\mathrm{eq}} \sim 3400$.
After that time, eventually the halo mass grew to
$M_{\mathrm{halo}} \simeq 100 M_{\mathrm{PBH}}$ until $z \sim 30$.

The inner radius of the dark matter halo is set to the innermost stable circular orbit 
of matter around the PBH,
which is
\begin{equation}
    \label{eq:2:3}
    r_{\mathrm{min}} = 6 G M_{\mathrm{PBH}},
\end{equation}
where $G$ is the gravitational constant.

The density profile of the dark matter halo is taken to be $\rho_{\mathrm{halo}} \propto r^{-9/4}$ in Newtonian Mechanics, as suggested in Refs.~\cite{Bertschinger:1985wig,Berezinsky:2013fxa,Adamek:2019gns,Serpico:2020ehh,Boudaud:2021irr}.
Moreover, the density profile is concretely expressed by
\begin{eqnarray}
    \label{eq:2:4}
    \rho_{\mathrm{halo}}(r) 
  &\sim& 8.701 \times 10^{3} \Big( \frac{1 + z}{1000} \Big)^{3} M_{\odot} \mathrm{pc}^{-3} \Big(\frac{r}{r_{\mathrm{max}}} \Big)^{ - n_{\mathrm{pro}}}, \\
  &\equiv& \rho_{0} \Big(\frac{r}{r_{\mathrm{min}}} \Big)^{ - n_{\mathrm{pro}}}, 
\end{eqnarray}
with $n_{\mathrm{pro}} = {9}/{4}$ for $\rho_{0} = 0.234 M_{\odot} \mathrm{pc}^{-3} (10^{2} M_{\mathrm{PBH}}/M_{\mathrm{halo}})^{3} (r_{\mathrm{max}}/r_{\mathrm{min}})^{n_{\mathrm{pro}}}$, which is derived from Refs.~\cite{Hertzberg:2020kpm,Oguri:2022fir}.

In this paper, we consider a halo composed of particle dark matter
with mass $m_{\chi} = 10^2 \text{-- \!}10^{4}\,\mathrm{GeV}$ surrounding a PBH
with mass $M_{\mathrm{PBH}} = 10^{18} \text{-- \!}10^{22}\, \mathrm{g}$.
The goal
is to calculate the flux resulting from the decay of dark matter in
the solar system.  The lifetime of the decaying particle dark matter
through $\chi \to \gamma + \gamma$ is strictly constrained by
observations.  In this study, we assume the lower bound of the
lifetime $\tau$ to be $10^{29}\, \mathrm{sec}$,
which is taken from the
observational data for $\mathcal{O}(1)\text{-- \!}\mathcal{O}(10^3)\,\mathrm{GeV}$
gamma-rays by the Fermi Gamma-ray Space Telescope~\cite{Foster:2022nva}.

We aim to calculate the average distance between PBHs under the assumption that they constitute a fraction of the CDM.
This fraction is defined as
\begin{equation}
    f_{\rm PBH} = \frac{\Omega_{\rm PBH}}{\Omega_{\rm CDM}},
\end{equation}
where $\Omega_{\rm PBH}$ ($\Omega_{\rm CDM}$) is the energy density parameter for the PBHs (CDM). Here the energy density parameter, $\Omega_i$, for a given component $i$ is defined as the ratio of its energy density, $\rho_{i}$, to the critical density, $\rho_{\rm crit}$, required for a flat Universe, i.e.,
\begin{equation}
    \Omega_{i} = \frac{\rho_{i}}{\rho_{\rm crit}}.
\end{equation}
This framework allows us to precisely quantify the contribution of PBHs to the overall dark matter content.
It is known that the density profile of dark matter around the solar system is $\rho_{\mathrm{DM}, \odot} = 0.4\text{-- \!}0.6 \,\mathrm{GeV/cm^{3}}$~\cite{deSalas:2020hbh}.
In this paper, we adopt $\rho_{\rm CDM, \odot} = 0.5 \,\mathrm{GeV/cm^{3}}$ for the CDM density around the solar system.
Assuming that the PBH density is given by $\rho_{\rm PBH, \odot} = f_{\rm PBH} \rho_{\rm CDM, \odot}$, we can determine the average distance $D$ between the PBHs
by using a relation, $D \sim (f_{\rm PBH} \rho_{\rm CDM, \odot}/ M_{\rm PBH})^{-1/3}$. In Table
\ref{tb:MPBH_and_Distance}, we show the average distance for each mass of a PBH at $f_{\rm PBH} = 1$.%
\footnote{From Eq.~\eqref{eq:2:1}, the fraction of CDM accounted for by PBHs is essentially $f_{\rm PBH} \leq 0.01$.
In this study, however, we hypothetically assume $f_{\rm PBH} = 1$ just for illustration purposes.}

\begin{table}[htbp]
    \centering
    \begin{tabular}{l||c|c|c}
        \hline
        $M_{\mathrm{PBH}}$ (g) & $10^{18}$ & $10^{20}$ & $10^{22}$ \\ \hline
        Distance $D$ (AU)         & $6.95$    & $32.2$    & $150$ \\ \hline
    \end{tabular}
    \caption{Values of the average distance for the PBHs with the mass $M_{\mathrm{PBH}}$, assuming $f_{\rm PBH} = 1$.}
    \label{tb:MPBH_and_Distance}
\end{table}

\label{sec:NW}

\section{New Calculations in General Relativity
}
\label{sec:GR}

\subsection{General relativistic model of dark matter halo around a primordial black hole}
To construct a spacetime model of a PBH
with a dark matter halo, we consider the general
static and spherically symmetric spacetime metric ansatz,
\begin{equation}
    \label{eq:3:1}
    g_{\mu \nu} dx^{\mu} dx^{\nu} = - \Big( 1 - \frac{2 \alpha(r)}{r} \Big) dt^{2} + \Big( 1 - \frac{2 G m(r)}{r} \Big)^{-1} dr^{2} + r^{2} d \Omega^{2},
\end{equation}
where $d \Omega^{2} = d \theta^{2} + \sin^{2}{\theta} d \phi^{2}$, and $x^{\mu} = (t, r, \theta, \phi)$ are time, areal radius, and angles on 2-sphere, respectively.
In our analysis, we focus on the static region, defined by the conditions
\begin{equation}
    \label{eq:3:2}
    m(r) < \frac{r}{2G}, \\
\end{equation}
\begin{equation}
    \label{eq:3:3}
    \alpha(r) < \frac{r}{2}.
\end{equation}

We assume a stress-energy tensor for a dark matter halo of the form
\begin{equation}
    \label{eq:3:4}
    T^{\mu}_{\nu} = \mathrm{diag}[- \rho(r), 0, P(r), P(r)],
\end{equation}
where $\rho(r)$ is the energy density, and $P(r)$ is the pressure uniformly applied in the direction tangent to each sphere.
Note that the stress-energy tensor~(\ref{eq:3:4})
is valid when the radial pressure is negligibly small compared to $\rho$ and $P$ 
(see, e.g., Refs.~\cite{Mahajan:2007vw,Igata:2022rcm}).

The energy density and pressure of
the dark matter halo are
related to the metric functions via the Einstein equations as
\begin{equation}
    \label{eq:3:5}
    \rho(r) = \frac{m'(r)}{4 \pi r^{2}},
\end{equation}
\begin{equation}
    \label{eq:3:6}
    P(r) = \frac{G m(r)}{2[r - 2 G m(r)]} \rho(r),
\end{equation}
where the prime denotes diﬀerentiation with respect to $r$.
The absence of radial pressure leads to
\begin{equation}
    \label{eq:3:7}
    \alpha'(r) = \frac{\alpha(r) - G m(r)}{r - 2 G m(r)}.
\end{equation}

We assume that a PBH
with mass $M_{\mathrm{PBH}}$ surrounded by a dark matter halo of the density $\rho_{\mathrm{halo}}(r)$ forms an isolated system.
In this model, the energy density is specified as
\begin{equation}
    \label{eq:3:8}
\rho(r) = 
\left\{ \,
    \begin{aligned}
    & 0 & (2 G M_{\mathrm{PBH}} < r \leq r_{\mathrm{min}}),\\
    & \rho_{\mathrm{halo}}(r) & (r_{\mathrm{min}} \leq r \leq r_{\mathrm{max}}),\\
    & 0 & (r \geq r_{\mathrm{max}}),
    \end{aligned}
\right.
\end{equation}
where $2 G M_{\mathrm{PBH}}$ is the Schwarzschild radius of the PBH.
Note that in both
the inner vacuum region $2 G M_{\mathrm{PBH}} < r \leq r_{\mathrm{min}}$ and the outer vacuum region $r \geq r_{\mathrm{max}}$, the metric 
reduces to the Schwarzschild according to
Birkhoﬀ's theorem.
Solving 
Eq.~\eqref{eq:3:5}
with this density profile yields the gravitational mass
\begin{equation}
    \label{eq:3:9}
m(r) = 
\left\{ \,
    \begin{aligned}
    & M_{\mathrm{PBH}} & (2 G M_{\mathrm{PBH}} < r \leq r_{\mathrm{min}}),\\
    & m_{\ast}(r) = M_{\mathrm{PBH}} + 4 \pi \int_{r_{\mathrm{min}}}^{r} r^{2} \rho_{\mathrm{halo}}(r) dr & (r_{\mathrm{min}} \leq r \leq r_{\mathrm{max}}),\\
    & M = M_{\mathrm{PBH}} + 4 \pi \int_{r_{\mathrm{min}}}^{r_{\mathrm{max}}} r^{2} \rho_{\mathrm{halo}}(r) dr & (r \geq r_{\mathrm{max}}),
    \end{aligned}
\right.
\end{equation}
where we have assumed the continuity of $m(r)$ at 
$r = r_{\mathrm{min}}$ and $r=r_{\mathrm{max}}$.
Thus, we can verify that $M$ is the total mass of the 
PBH and the dark matter halo.
Furthermore, we obtain $\alpha(r)$ by solving 
Eq.~\eqref{eq:3:7}
as
\begin{equation}
    \label{eq:3:10}
\alpha(r) = 
\left\{ \,
    \begin{aligned}
    & \frac{r}{2} - \frac{C^{2}}{2}(r - 2 G M_{\mathrm{PBH}}) & (2 G M_{\mathrm{PBH}} < r \leq r_{\mathrm{min}}),\\
    & \alpha_{\ast}(r) = \frac{r}{2} - \frac{r_{\mathrm{max}} - 2 G M}{2} \exp{ \Big( \int_{r_{\mathrm{max}}}^{r} \frac{dr}{r - 2 G m_{\ast}(r)} \Big)}& (r_{\mathrm{min}} \leq r \leq r_{\mathrm{max}}),\\
    & G M & (r \geq r_{\mathrm{max}}),
    \end{aligned}
\right.
\end{equation}
with
\begin{equation}
    \label{eq:3:11}
    C^{2} = \frac{r_{\mathrm{max}} - 2G M}{r_{\mathrm{min}} - 2 G M_{\mathrm{PBH}}} \exp{ \Big( - \int_{r_{\mathrm{min}}}^{r_{\mathrm{max}}} \frac{dr}{r - 2 G m_{\ast}(r)} \Big)},
\end{equation}
where we have chosen the gauge in 
Eq.~\eqref{eq:3:10}
so that the metric appears in the standard Schwarzschild coordinates in the outer vacuum region and have assumed the continuity of $\alpha(r)$ at 
$r = r_{\mathrm{min}}$ and $r=r_{\mathrm{max}}$.
Note that when $C^2$ is evaluated using the parameters specified in Sec.~\ref{sec:NW}, the resulting value is nearly equal to unity (i.e., $C^2\sim1$).
Accordingly, the metric in the vacuum region reduces to
\begin{equation}
    \label{eq:3:12}
g_{\mu \nu} dx^{\mu} dx^{\nu} = 
\left\{ \,
    \begin{aligned}
    & - C^{2} \Big(1 - \frac{2 G M_{\mathrm{PBH}}}{r} \Big) dt^{2} + \Big(1 - \frac{2 G M_{\mathrm{PBH}}}{r} \Big)^{-1} dr^{2} + r^{2} d \Omega^{2} & (2 G M_{\mathrm{PBH}} < r \leq r_{\mathrm{min}}),\\
    & - \Big(1 - \frac{2 G M}{r} \Big) dt^{2} + \Big(1 - \frac{2 G M}{r} \Big)^{-1} dr^{2} + r^{2} d \Omega^{2} & (r \geq r_{\mathrm{max}}),
    \end{aligned}
\right.
\end{equation}
where $t$ coincides with the Schwarzschild time in the outer region but not in the inner region.
In summary, this model describes a spacetime with a Schwarzschild black hole of mass $M_{\mathrm{PBH}}$ at the center, a static self-gravitating dark matter halo in a bounded region around it, and the Schwarzschild vacuum outside the halo, thus representing an asymptotically flat 
PBH surrounded by a dark matter halo.

Suppose the halo consists of a collisionless cluster of self-gravitating dark matter particles in geodesic circular motion.
This picture is known as the Einstein cluster \cite{Einstein:1939ms} and is compatible with the stress-energy 
tensor~\eqref{eq:3:4}.
Let $n(r)$ be the number density profile.
Then, $\rho_{\mathrm{halo}}(r)$ and the tangential pressure of the halo $P_{\mathrm{halo}}(r)$ take the form
\begin{equation}
    \label{eq:3:13}
    \rho_{\mathrm{halo}}(r) = m_{\chi} n(r) \frac{r - 2 G m_{\ast}(r)}{r - 3 G m_{\ast}(r)} \geq 0,
\end{equation}
\begin{equation}
    \label{eq:3:14}
    P_{\mathrm{halo}}(r) = m_{\chi} n(r) \frac{G m_{\ast}(r)}{2[r - 3 G m_{\ast}(r)]} \geq 0,
\end{equation}
These expressions imply
that $m_{\ast}(r)$ is a non-decreasing function in $r$ and the range of $r$ is more restricted than 
Eq.~\eqref{eq:3:2},
\begin{equation}
    \label{eq:3:15}
    m'_{\ast} \geq 0,
\end{equation}
\begin{equation}
    \label{eq:3:16}
    r > 3 G m_{\ast}(r).
\end{equation}

\subsection{Dynamics of free photons}
We consider the dynamics of photons in
the spacetime constructed in the previous section, 
under the assumption
that local interactions between photons and matter are negligible.
Due to the spherical symmetry of the background, we can choose coordinates in which a free photon remains confined to
the equatorial plane $ \theta= \pi/2$.
In these coordinates, the Lagrangian of a free photon reduces to
\begin{equation}
    \label{eq:4:1}
    \mathscr{L} = \frac{1}{2} \Big[- \Big(1 - \frac{2 \alpha(r)}{r} \Big) \dot{t}^{2} + \Big(1 - \frac{2 G m(r)}{r} \Big)^{-1} \dot{r}^{2} + r^{2} \dot{\phi}^{2} \Big],
\end{equation}
where the dot denotes diﬀerentiation with respect to an aﬃne parameter $\lambda$.
Since $t$ and $\phi$ are cyclic coordinates,
the Euler-Lagrange equations for $t$ and $\phi$ lead to the conservation of their conjugate momenta
\begin{equation}
    \label{eq:4:2}
    \frac{\partial{\mathscr{L}}}{\partial \dot{t}} = - \Big( 1 - \frac{2 \alpha(r)}{r} \Big) \dot{t} = - E,
\end{equation}
\begin{equation}
    \label{eq:4:3}
    \frac{\partial{\mathscr{L}}}{\partial \dot{\phi}} = r^{2} \dot{\phi} = L,
\end{equation}
where $E$ and $L$ are conserved energy and angular momentum, respectively.
With these constants, the null condition, $\mathscr{L} = 0$, gives the equation of radial motion
\begin{equation}
    \label{eq:4:4}
    \frac{\dot{r}^{2}}{2} + U(r) = 0,
\end{equation}
\begin{equation}
    \label{eq:4:5}
    U(r) = \frac{1}{2} \Big( 1 - \frac{2 G m(r)}{r} \Big) \Big( \frac{b^{2}}{r^{2}} - \frac{r}{r - 2 \alpha(r)} \Big),
\end{equation}
where from now on, the dot denotes diﬀerentiation with respect to the rescaled 
aﬃne parameter $E \lambda \to \lambda$, and $b$ is the impact parameter defined as
\begin{equation}
    \label{eq:4:6}
    b = \frac{L}{E}.
\end{equation}
By diﬀerentiating Eq.~\eqref{eq:4:4},
we can obtain the Euler-Lagrange equation for the radial variable,
\begin{equation}
    \label{eq:4:7}
    \ddot{r} + U'(r) = 0,
\end{equation}
where the prime denotes diﬀerentiation with respect to $r$.

Now we focus on photon spheres.
By imposing the conditions for a photon in circular motion, $\dot{r} = 0$ and $\ddot{r} = 0$, on Eqs.~\eqref{eq:4:4} and \eqref{eq:4:7},
we obtain
\begin{equation}
    \label{eq:4:8}
    U(r) = 0,
\end{equation}
\begin{equation}
    \label{eq:4:9}
    U'(r) = 0.
\end{equation}
Solving these equations simultaneously yields
\begin{equation}
    \label{eq:4:10}
    r = 3 G m(r),
\end{equation}
\begin{equation}
    \label{eq:4:11}
    b^{2} = \frac{r^{3}}{r - 2 \alpha(r)}.
\end{equation}
Equation~(\ref{eq:4:10})
determines the radii of photon spheres, and Eq.~\eqref{eq:4:11}
gives the impact parameters for photons moving on them.
If a photon sphere exists in the inner vacuum region,
then the photon sphere radius is
\begin{equation}
    \label{eq:4:12}
    r = 3 G M_{\mathrm{PBH}}.
\end{equation}
Substituting this result into Eq.~\eqref{eq:4:11} then leads to the expression for the critical impact parameter,
\begin{equation}
    \label{eq:4:13}
    b_{\mathrm{c}}^2
    = \frac{27 G^{2} M_{\mathrm{PBH}}^{2}}{C^{2}},
\end{equation}
where Eq.~(\ref{eq:3:10}) is used.
This result implies that the presence of the dark matter halo induces a deviation in $b_{\mathrm{c}}$ relative to the vacuum case. However, since $C^2\sim 1$ in the current configuration, we obtain $b_{\mathrm{c}}
\sim 3 \sqrt{3} G M_{\mathrm{PBH}} \sim 5.1 G M_{\mathrm{PBH}}$, which is nearly identical to the value for the Schwarzschild spacetime.
In other words, photons moving in unstable circular orbits remain at $r=3GM_{\mathrm{PBH}}$, while those escaping from the photon sphere form a light ring with an apparent size corresponding to an impact parameter $b_{\mathrm{c}} \sim  5.1 G M_{\mathrm{PBH}}$.
On the other hand, there is no photon sphere in the halo region because Eq.~\eqref{eq:4:10}
is incompatible with the condition~\eqref{eq:3:16}
for the Einstein clusters.

\subsection{How is a dressed PBH observed by using photons?
}
We derive an expression for the observed radiance resulting from isotropic and monochromatic radiation emitted
by the dark matter decay in the halo~\cite{Bambi:2013nla}.
Let $\langle u^{a}_{\mathrm{halo}} \rangle$ be the averaged velocity of the halo,
\begin{equation}
    \label{eq:5:1}
    \langle u^{a}_{\mathrm{halo}} \rangle = \frac{1}{\sqrt{1 - 2 \alpha_{\ast}(r)/r}} (\partial/\partial t)^{a},
\end{equation}
which defines the halo's rest frame.
Let $\xi^{a}$ be the timelike Killing vector,
\begin{equation}
    \label{eq:5:2}
    \xi^{a} = (\partial/\partial t)^{a}.
\end{equation}
Let $j_{\nu_{\mathrm{emit}}}$ be the spectral intensity per unit volume (i.e., the radiant energy density per unit time, per unit solid angle, and per unit frequency) measured in the rest frame of the halo, where $\nu_{\mathrm{emit}}$ is the
photon frequency 
at the emission point in that frame.
Assume that a photon, emitted in the halo with frequency $\nu_{\mathrm{emit}}$,
propagates along a null geodesic $\mathcal{C}$ and is observed at frequency $\nu_{\mathrm{obs}}$ by a distant observer.
Let $k^{a}$ be the tangent vector of $\mathcal{C}$.
We define the redshift factor and rewrite it by the geometrical quantity as
\begin{equation}
    \label{eq:5:3}
    g = \frac{\nu_{\mathrm{obs}}}{\nu_{\mathrm{emit}}} = \frac{(k_{a} \xi^{a})|_{\mathrm{obs}}}{(k_{a} \langle u^{a}_{\mathrm{halo}} \rangle)|_{\mathrm{emit}}} = \sqrt{1 - \frac{2 \alpha_{\ast}(r)}{r}},
\end{equation}
where the symbols “obs” and “emit” mean evaluating the results at the observation point and the emission point, respectively.
Since the lower limit of $g$ occurs when $r = r_{\mathrm{min}}$, the range of $g$ is $g \in (0.816,1.00)$.

The spectral radiance (or sometimes called the specific intensity) of $\nu_{\mathrm{obs}}$ is given by
\begin{equation}
    \label{eq:5:4}
    I_{\nu_{\mathrm{obs}}}(X,Y) = \int_{\tilde{\mathcal{C}}} g^{3} j_{\nu_{\mathrm{emit}}} dl.
\end{equation}
Here, $(X, Y)$ are the coordinates on the celestial sphere, and we have applied the reciprocity theorem ~\cite{Etherington_1933} (or equivalently, Liouville’s theorem).
The spacelike curve $\tilde{\mathcal{C}}$ represents the projection of $\mathcal{C}$ onto a constant-$t$ hypersurface, which is orthogonal to the halo’s average velocity field.
The infinitesimal proper length along $\tilde{\mathcal{C}}$ is expressed as
\begin{equation}
    \label{eq:5:5}
    dl = - k_{a} \langle u^{a}_{\mathrm{halo}} \rangle d \lambda = \frac{|dr|}{g \sqrt{- 2 U(r)}},
\end{equation}
where we have used 
Eq.~\eqref{eq:4:4}~\cite{Bambi:2013nla}.
Furthermore, we assume that the radiation is isotropic in the rest frame and monochromatic 
at the rest frame frequency $\nu_{\gamma}$ and 
is produced by
dark matter decay with the lifetime $\tau$, 
\begin{equation}
    \label{eq:5:6}
    j_{\nu_{\mathrm{emit}}} = \frac{\rho_{\mathrm{halo}}(r)}{4 \pi \tau} \delta(\nu_{\mathrm{emit}} - \nu_{\gamma}).
\end{equation}
Then, the spectral radiance $I_{\nu_{\mathrm{obs}}}$ is written as
\begin{equation}
    \label{eq:5:7}
    I_{\nu_{\mathrm{obs}}}(X,Y) = \int_{\tilde{\mathcal{C}}} \frac{1}{4 \pi \tau} \frac{g^{3} \rho_{\mathrm{halo}}(r)}{\sqrt{- 2 U(r)}} \frac{\delta(\nu_{\mathrm{emit}} - \nu_{\gamma})}{g} |dr|.
\end{equation}
Integrating the spectral radiance
over the frequency domain of $\nu_{\mathrm{obs}}$ yields the radiance on the celestial sphere of the observer,
\begin{equation}
    \label{eq:5:8}
    I(X,Y) = \int d \nu_{\mathrm{obs}} I_{\nu_{\mathrm{obs}}}(X,Y) = \frac{1}{4 \pi \tau} \int_{\tilde{\mathcal{C}}} \frac{g^{3} \rho_{\mathrm{halo}}(r)}{\sqrt{- 2 U(r)}} |dr|,
\end{equation}
where $b^{2} = X^{2} + Y^{2}$.

\begin{figure}
\centering
\includegraphics[width=160mm]{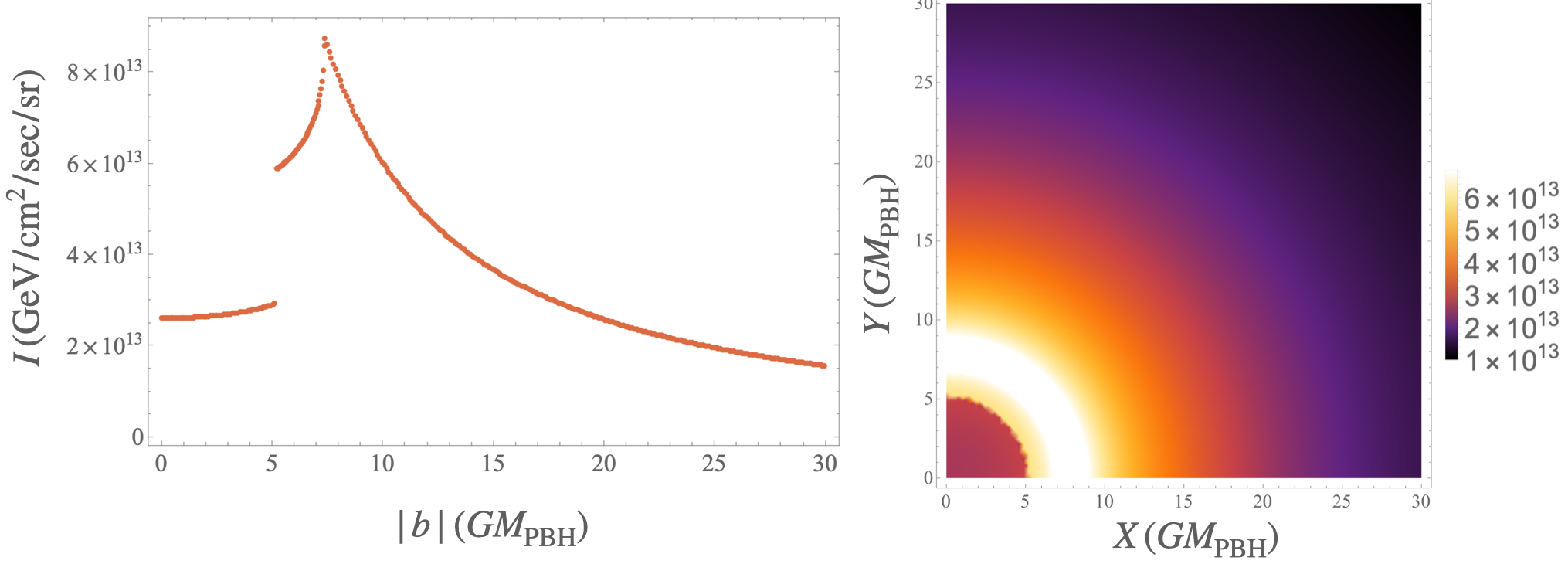} 
\caption{
Left: Radiance from decaying dark matter around the PBH ($M_{\mathrm{PBH}} = 10^{18} \,\mathrm{g}$), evaluated at the outer boundary of the halo and expressed as a function of the impact parameter $b$.
For $|b|\leq b_{\mathrm{c}}\sim 5.1 G M_{\mathrm{PBH}}$,
the flux from the decaying dark matter only on the observer's side can be observed.
For $|b|\geq b_{\mathrm{c}}$, the radiance
from all directions can be observed, resulting in a more than twofold increase.
The maximum radiance
occurs at $b_{\mathrm{in}} \sim 7.3 G M_{\mathrm{PBH}}$, which corresponds to the impact parameter at the inner edge of the dark matter halo, $r_{\mathrm{min}}$. 
Right: The same radiance distribution is depicted as density contours on the $X$-$Y$ plane (assuming $b^2=X^{2}+Y^{2}$).
}
\label{ring}
\end{figure}

We examine how bright a PBH halo
appears on the celestial sphere.
Figure~\ref{ring}
shows the results of numerical calculations for $M_{\mathrm{PBH}} = 10^{18} \,\mathrm{g}$.
It is important to note that these results do not take into account the distance to the observer and represent the radiance around the dark matter halo.
If $|b|\leq b_{\mathrm{c}}$,
the flux from the decay of dark matter on the observer's side can be observed.

On the other hand, if $|b| > b_{\mathrm{c}}$,
scattering due to the effective potential occurs, and the light rays traversing the halo integrate the intensity contributions both as they approach and as they recede from the point of closest approach.
Hence, the radiance distribution has a gap at 
$|b| = b_{\mathrm{c}}$.
For $|b|\geq b_{\mathrm{c}}$,
the radiance has a peak at $|b| = b_{\mathrm{in}}$,
where
\begin{equation}
    \label{eq:5:9}
    b_{\mathrm{in}} = \frac{r_{\mathrm{min}}}{C \sqrt{1 - 2 G M_{\mathrm{PBH}}/r_{\mathrm{min}}}},
\end{equation}
which is obtained by $U(r_{\rm min})=0$ with $|b|=b_{\rm in}$.
Here, the result is numerically $b_{\mathrm{in}} \sim 7.3 G M_{\mathrm{PBH}}$, using Eq.~\eqref{eq:2:3}.
This means that the bright ring does not appear on the shadow edge but a bit away from it.
The scale of the structure corresponds to the scale of the inner boundary of the halo.
It can be seen that the radiance due to decay is maximal at
$r_{\mathrm{min}}$, where the dark matter halo attains its maximum density.
For $b_{\mathrm{in}} \leq |b| \leq b_{\mathrm{out}}$,
$I(X, Y)$ 
decreases monotonically with increasing $|b|$
and vanishes at $|b|= b_{\mathrm{out}}$,
where
\begin{equation}
    \label{eq:5:10}
    b_{\mathrm{out}} = \frac{r_{\mathrm{max}}}{\sqrt{1 - 2G M_{\mathrm{PBH}}/r_{\mathrm{max}}}},
\end{equation}
which is obtained by $U(r_{\mathrm{max}}) = 0$ with $|b|=b_{\rm out}$.
For $|b| \geq b_{\mathrm{out}}$,
$I(X, Y)$ vanishes because no light ray passes through the halo.
These behaviors must be typical in this model.

\begin{figure}
\centering
\includegraphics[width=130mm]{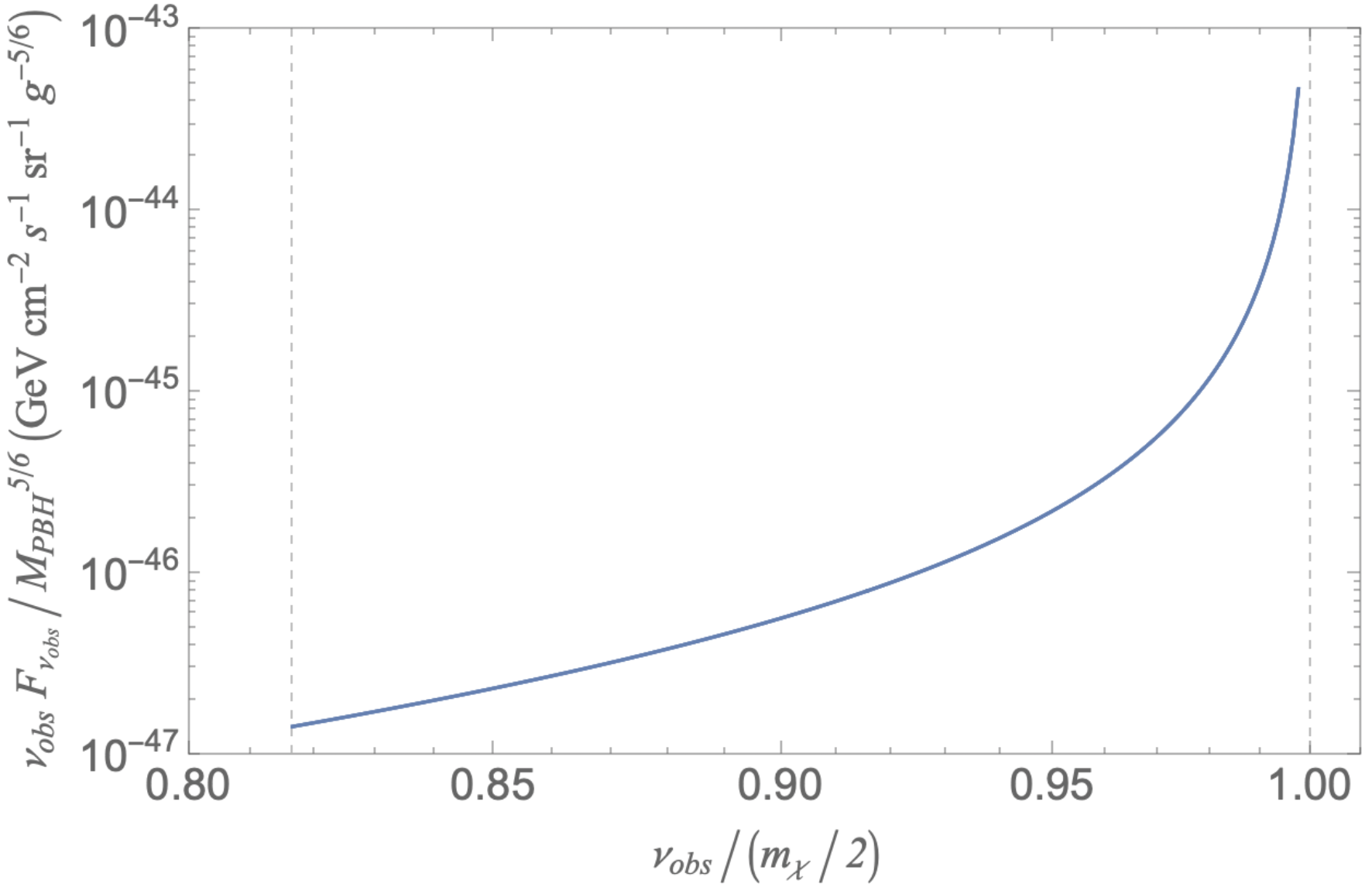} 
\caption{
  Mass-independent spectrum at $f_{\rm PBH} = 1$.
  The $x$-axis represents the energy of the photon normalized by $m_{\chi}/2$, where $m_{\chi}$ is the mass of the dark matter particle.
  This normalization is chosen due to the decay process $\chi \to \gamma + \gamma$.
  The flux is restricted to the range $g \in (0.816,1.00)$, and it is cut off beyond this interval.
  Note that, unlike Fig.~\ref{ring}, 
  which investigates the relationship with the impact parameter $b$, we focus here on $g$, representing the distance $r$ from the center of the PBH where the decay occurs.}
\label{nuF}
\end{figure}

Finally, we calculate the flux received by a distant observer at a distance $D$, which 
is expressed by
\begin{equation}
    \label{eq:5:11}
    \nu_{\mathrm{obs}} F_{\nu_{\mathrm{obs}}} = \int_{\tilde{\mathcal{C}}}
 dr dX dY \frac{\nu_{\mathrm{obs}}}{4 \pi D^{2}} \frac{1}{4 \pi \tau} \frac{1}{2} \frac{g^{3} \rho_{\mathrm{halo}}(r)}{\sqrt{- 2 U(r)}} \delta(\nu_{\mathrm{emit}} - \nu_{\gamma}).
\end{equation}
Due to the decay of dark matter ($\chi \to \gamma + \gamma$), we take the flux to be half in Eq.~\eqref{eq:5:11}.
This flux is approximately proportional to $f_{\rm PBH}^{2/3} M_{\mathrm{PBH}}^{5/6}$.%
\footnote{That is
  because the distance $D$ is proportional to
  $D \propto f_{\rm PBH}^{- 1/3} M_{\rm PBH}^{1/3}$. If we fix the distance without
  depending on $M_{\rm PBH}$, we obtain
  $\nu_{\mathrm{obs}} F_{\nu_{\mathrm{obs}}} \propto M_{\rm
    PBH}^{3/2}$. 
}
For details, see Appendix~\ref{sec:App2}.
Moreover,
the transformation used to calculate the delta function shows
that it does not depend on the mass of dark matter.  Therefore, by
calculating $\nu_{\mathrm{obs}} F_{\nu_{\mathrm{obs}}}/M_{\mathrm{PBH}}^{5/6}$, we obtain
the mass-independent spectrum.  By considering
the gravitational redshifts, we also obtained the broadened spectrum by the general relativistic effect.
In Fig.~\ref{nuF}, we plot the spectrum at $f_{\rm PBH} = 1$.

\begin{figure}
\centering
\includegraphics[width=130mm]{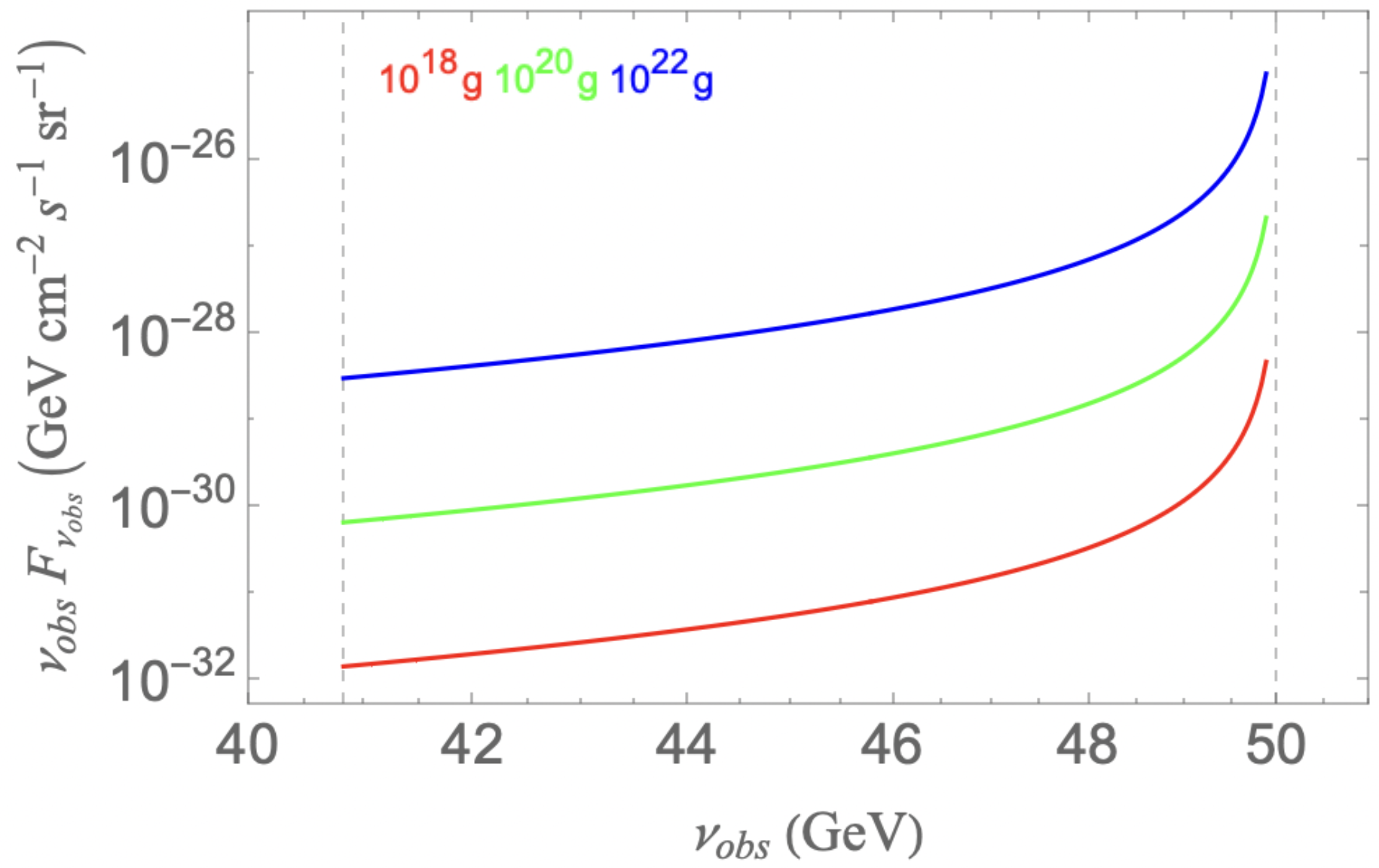} 
\caption{Photon flux for $m_{\chi} = 100 \,\mathrm{GeV}$
with $M_{\mathrm{PBH}} = 10^{18} \,\mathrm{g}$, $10^{20} \,\mathrm{g}$, and $10^{22}\, \mathrm{g}$
 from the bottom to the top at $f_{\rm PBH} = 1$. The $x$-axis represents the photon energy in GeV.}
\label{18-20-22-100}
\end{figure}

We now consider some specific cases for the mass of the PBHs.
In Fig.~\ref{18-20-22-100},
we plot 
the photon flux arising from particle dark matter with mass $m_{\chi} = 100 \,\mathrm{GeV}$ for three PBH masses:
$M_{\mathrm{PBH}} = 10^{18}$~g, $10^{20}$~g, and $10^{22}$~g.
Due to the effect of the gravitational redshifts, we obtain the broadened
spectrum. We also confirm that the flux increases with 
PBH mass, following the scaling of $\propto M_{\mathrm{PBH}}^{5/6}$.

It is important to note that, for simplicity, we have considered
the flux emitted only
from the nearest single system of a dressed PBH.
In fact, we have to
integrate the galactic or extra-galactic photon spectra from all
contributing PBHs, which can be compared with data taken by the
current and future observations~\cite{CTAConsortium:2010umy}. 
These topics will be studied in a separate paper~\cite{Suzuki2025Part2}.

\section{Conclusion}
In this paper, we have investigated general relativistic effects
on the photon spectrum emitted by the decay or annihilation 
within the
halo of particle dark matter surrounding a primordial black hole. Due
to the influence of strong gravitational fields near the black hole,
the line spectrum undergoes broadening as a consequence of
gravitational redshifts, resulting in a broad spectrum.

In the case where particle dark matter decays and emits line photons,
we have analytically derived that the flux follows the relation $\nu_{\mathrm{obs}} F_{\nu_{\mathrm{obs}}} \propto f_{\rm PBH}^{2/3} M_{\rm PBH}^{5/6}$ 
with $M_{\rm PBH}$ being the mass
of the primordial black hole. Additionally, as demonstrated in
Appendix~\ref{sec:App1}, the flux is proportional to $\nu_{\mathrm{obs}} F_{\nu_{\mathrm{obs}}}
\propto f_{\rm PBH}^{2/3} M_{\rm PBH}^{-2/3} m_{\chi}^{-1}$
with mass $m_{\chi}$ of annihilating particle dark matter.

The detection of such a distinctive photon spectrum by future
observations of electromagnetic waves would provide crucial evidence
supporting the mixed dark matter scenario involving both PBHs and
particle dark matter. We expect that this type of study, in turn,
could offer significant insights into the fundamental nature of dark
matter.

\acknowledgments
We thank 
Nozomu Tominaga, Tomoya Takiwaki, and Yohsuke Takamori
for their fruitful
discussions. This work was supported in part by JSPS KAKENHI Grants 
No.~JP22K03611, No.~JP23KK0048, No.~JP24H00183 (T.I.), No.~JP24K07027 (K.K. and T.H.) and No.~JP23KF0289 (K.K.), and MEXT KAKENHI Grants No.~JP24H01825 (K.K.) and No.~JP20H05853 (T.H.).

\appendix
\section{Case for Annihilations}
Here, we consider the case for annihilating particle dark matter $\chi$ through the process, $\chi + \chi \to \gamma + \gamma$ inside the halo formed around a primordial black hole.
In this case, Eq.~\eqref{eq:5:6} for the decaying particle dark matter is replaced by
\begin{equation}
    j_{\nu_{\mathrm{emit}}} = \frac{\rho_{\mathrm{halo}}^{2}(r) \langle \sigma_{\mathrm{ann}} v \rangle}{4 \pi m_{\chi}} \delta(\nu_{\mathrm{emit}} - \nu_{\gamma}),
\end{equation}
where $\langle \sigma_{\mathrm{ann}} v \rangle$ is the thermal-averaged annihilation cross section.
We adopt the concrete value of $\langle \sigma_{\mathrm{ann}} v \rangle = 3 \times 10^{-26} \mathrm{cm^{3} s^{-1}}$ that can explain the current relic abundance of dark matter given in Ref.~\cite{Steigman:2012nb}.
The calculation of Eq.~\eqref{eq:5:11} is then carried out with this modification.
Then, we can write
\begin{equation}
    \label{eq:A:2}
    \nu_{\mathrm{obs}} F_{\nu_{\mathrm{obs}}} = \int_{\tilde{\mathcal{C}}} dr dX dY \frac{\nu_{\mathrm{obs}}}{4 \pi D^{2}} \frac{\langle \sigma_{\mathrm{ann}} v \rangle}{4 \pi m_{\chi}} \frac{g^{3} \rho_{\mathrm{halo}}^{2}(r)}{\sqrt{- 2 U(r)}} \delta(\nu_{\mathrm{emit}} - \nu_{\gamma}).
\end{equation}
By using Eqs.~\eqref{eq:2:1}--\eqref{eq:2:4},
analytically we
obtain the approximate form of the flux to be
$\nu_{\mathrm{obs}} F_{\nu_{\mathrm{obs}}} \propto f_{\rm PBH}^{2/3} M_{\mathrm{PBH}}^{-2/3} m_{\chi}^{-1}$.
Note that the distance $D$ is proportional to 
$(f_{\rm PBH} \rho_{\rm CDM, \odot}/M_{\rm PBH})^{-1/3} \propto f_{\rm PBH}^{- 1/3} M_{\rm PBH}^{1/3}$
with being $\rho_{\rm PBH, \odot}$ the mass density of the PBHs.  Thus, if the
distance $D$ is fixed, it is confirmed that the result is independent
of the mass of the primordial black hole.

\begin{figure}
\centering
\includegraphics[width=130mm]{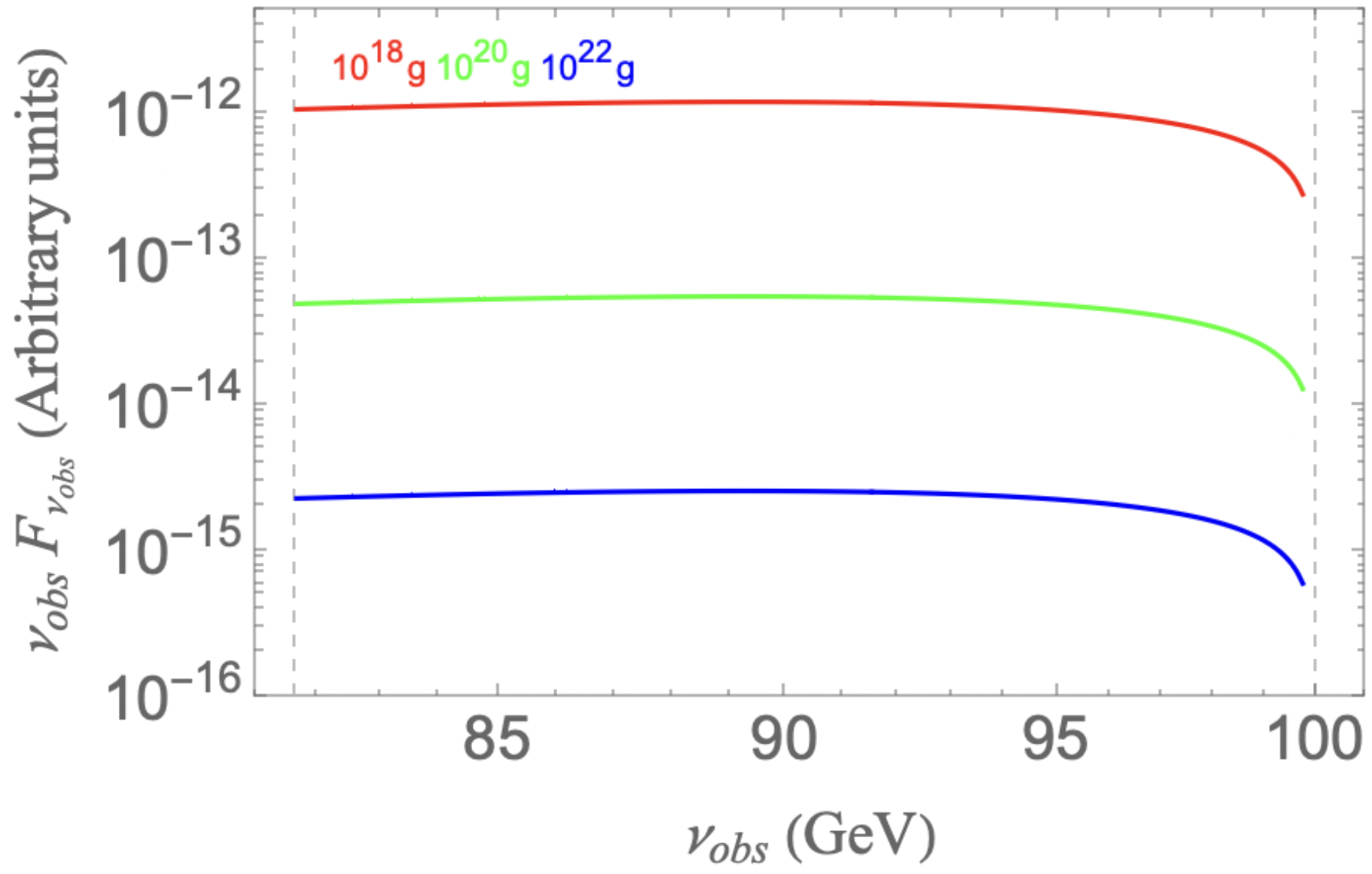} 
\caption{Numerical evaluations of the spectrum given in Eq.~\eqref{eq:A:2} for $m_{\chi} = 100\, \mathrm{GeV}$ in three cases for the masses of primordial black holes ($M_{\mathrm{PBH}}=10^{18}$, $10^{20}$, and $10^{22}$~g from top to bottom) at $f_{\rm PBH} = 1$.
The smaller the mass of the primordial black hole, the larger the flux. Here, the $y$-axis is plotted in arbitrary units.}
\label{arb100G}
\end{figure}

\begin{figure}
\centering
\includegraphics[width=130mm]{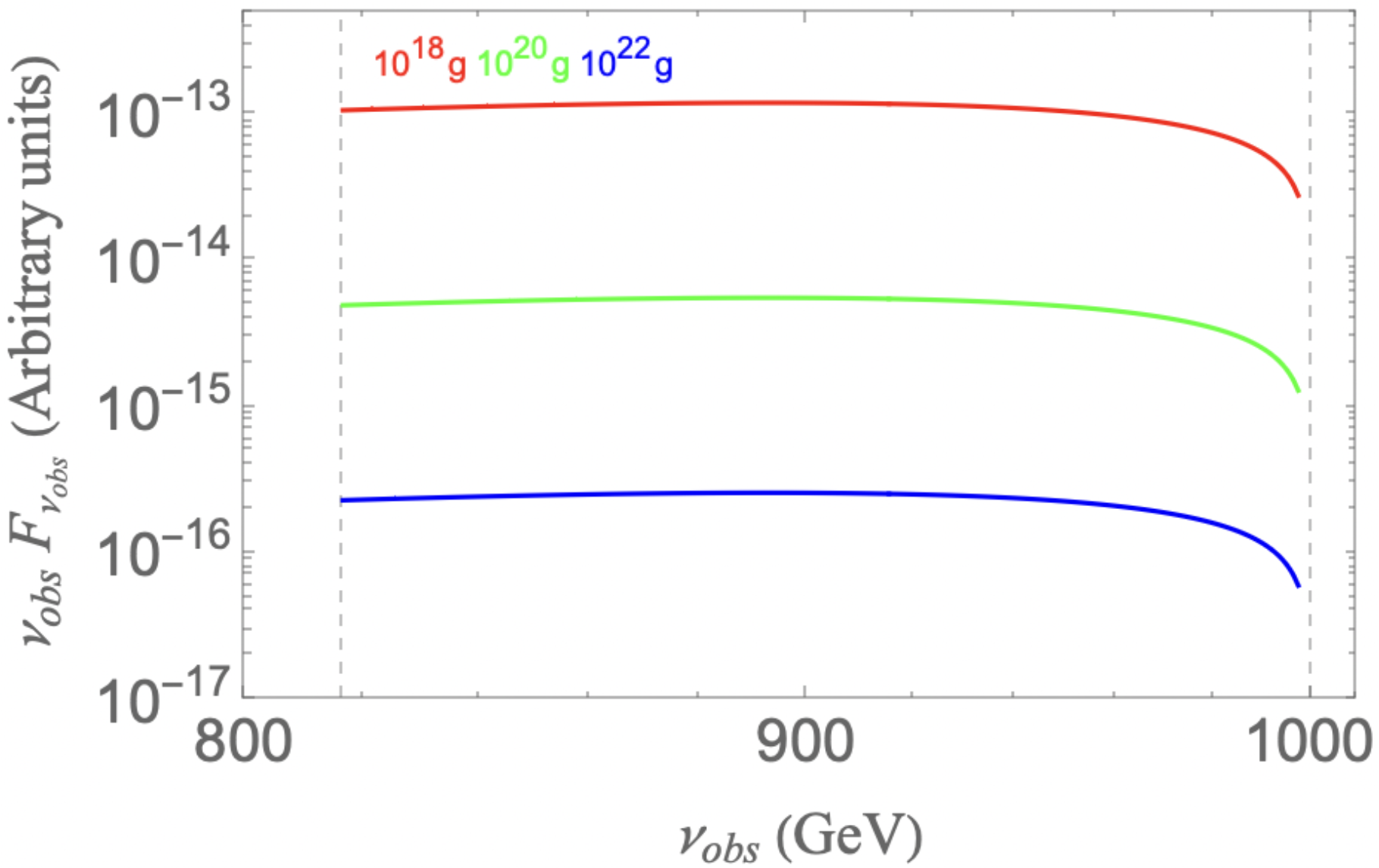} 
\caption{The same as Fig.~\ref{arb100G}, except for $m_{\chi} = 1\, \mathrm{TeV}$ at $f_{\rm PBH} = 1$.
}
\label{arb1T}
\end{figure}

\begin{figure}
\centering
\includegraphics[width=130mm]{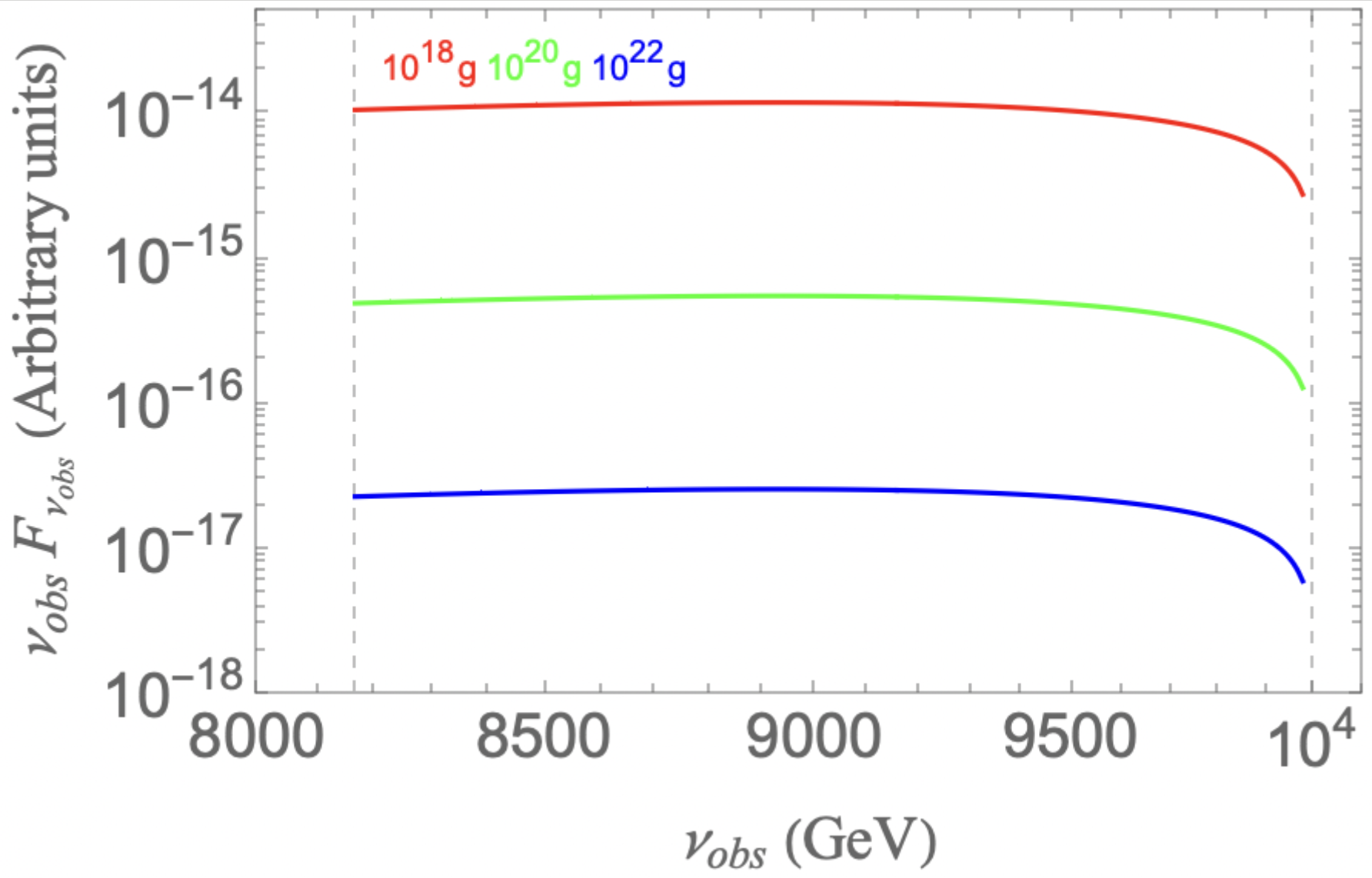} 
\caption{The same as Fig.~\ref{arb100G}, except for $m_{\chi} = 10 \,\mathrm{TeV}$ at $f_{\rm PBH} = 1$.
}
\label{arb10T}
\end{figure}

Next, we calculate some examples by using Eq.~\eqref{eq:A:2}.
For a fixed value of the mass of the particle dark matter to be $m_{\chi} = 100 \,\mathrm{GeV}$, we evaluate Eq.~\eqref{eq:A:2} in three cases for the masses of the primordial black holes ($M_{\mathrm{PBH}}=10^{18}$~g, $10^{20}$~g, and $10^{22}$~g), as shown in Fig.~\ref{arb100G}.
In these figures, note that the $y$-axis is shown in
arbitrary units.
Similarly, the results for $m_{\chi} = 1\, \mathrm{TeV}$
and $10\, \mathrm{TeV}$
are presented in Figs.~\ref{arb1T} and \ref{arb10T}, respectively.

There is some debate regarding the density profile.  According to Refs.~\cite{Carr:2020mqm,Kadota:2022cij,Eroshenko:2024dtb}, the dark matter halo around the current primordial black hole continues annihilating, leading to a decrease in the density profile. The temporal change in
this density profile has a significant impact on the flux. In
Appendix~\ref{sec:App2}, we consider general cases for the exponent
of the power-law density profile.

\label{sec:App1}

\section{General Exponent of Power-Law Density Profile}
We have discussed the photon spectrum by simply assuming
$n_{\mathrm{pro}} = 9/4$ in Eq.~\eqref{eq:2:4}.
However, it is nontrivial to consider how a change
in the exponent affects the flux, which has been discussed in some works~\cite{Boudaud:2021irr,Eroshenko:2024dtb}.
Finally, we derive the analytical dependences of the flux on the PBH mass $M_{\mathrm{PBH}}$ for both decaying and annihilating dark matter.
Except for the generalized versions of Eq.~\eqref{eq:2:4}, we use the same equations as those used
in Sec.~\ref{sec:NW}.
Note that in this model, the distance $D$ is proportional to $(f_{\rm PBH} \rho_{\rm CDM, \odot}/M_{\rm PBH})^{-1/3} \propto f_{\rm PBH}^{- 1/3} M_{\rm PBH}^{1/3}$.
Given that the distances $r$, $X$, and $Y$ around PBH are normalized by the mass $M_{\mathrm{PBH}}$, the density near $r_{\mathrm{max}}$ becomes sufficiently small and barely contributes to the integral.
Therefore, the integral converges at a radius much smaller than $r_{\mathrm{max}}$.
From Eqs.~\eqref{eq:2:1}--\eqref{eq:2:4}, we find that $\rho_{\mathrm{halo}}(r)$ scales as $M_{\mathrm{PBH}}^{- \frac{2}{3} n_{\mathrm{pro}}}$, and using Eqs.~\eqref{eq:4:5} and \eqref{eq:3:10}, $U(r)$ is proportional to $M_{\mathrm{PBH}}^{0}$.

From the same calculations as those in Sec.~\ref{sec:GR}, we can calculate that the flux produced by the decaying dark matter (Eq.~\eqref{eq:5:11}) is approximately represented by
\begin{equation}
    \nu_{\mathrm{obs}} F_{\nu_{\mathrm{obs}}} \propto f_{\rm PBH}^{\frac{2}{3}} M_{\mathrm{PBH}}^{- \frac{2}{3}(1 + n_{\mathrm{pro}}) + 3}
\end{equation}
for decaying dark matter. In fact, substituting $n_{\mathrm{pro}} = 9/4$, we obtain $\nu_{\mathrm{obs}} F_{\nu_{\mathrm{obs}}} \propto f_{\rm PBH}^{2/3} M_{\mathrm{PBH}}^{5/6} $,
which is consistent with the results in Sec.~\ref{sec:GR}.

On the other hand, Eq.~\eqref{eq:A:2} approximately gives
\begin{equation}
    \nu_{\mathrm{obs}} F_{\nu_{\mathrm{obs}}} \propto f_{\rm PBH}^{\frac{2}{3}} M_{\mathrm{PBH}}^{- \frac{2}{3}(1 + 2 n_{\mathrm{pro}}) + 3},
\end{equation}
for annihilating dark matter.

\label{sec:App2}

\bibliographystyle{apsrev4-2} 
\bibliography{main} 

\end{document}